\documentclass[runningheads]{llncs}
\usepackage{graphicx}
\usepackage{amsmath, bm}
\usepackage{braket}

\begin{document}

\title{Fast Swapping in a Quantum Multiplier Modelled as a Queuing Network}

\author{Evan E. Dobbs\inst{1} \and
Robert Basmadjian\inst{2} \and \\
Alexandru Paler\inst{1,3} \and
Joseph S. Friedman\inst{1}}
\authorrunning{E. Dobbs et al.}
\institute{University of Texas at Dallas, Richardson, TX 75080, USA
\and
Clausthal University of Technology, 38678 Clausthal-Zellerfeld, Germany
\and
Transilvania University, 500036 Brașov, România
}

\maketitle

\begin{abstract}
Predicting the optimum SWAP depth of a quantum circuit is useful because it informs the compiler about the amount of necessary optimization. Fast prediction methods will prove essential to the compilation of practical quantum circuits. In this paper, we propose that quantum circuits can be modeled as queuing networks, enabling efficient extraction of the parallelism and duration of SWAP circuits. To provide preliminary substantiation of this approach, we compile a quantum multiplier circuit and use a queuing network model to accurately determine the quantum circuit parallelism and duration. Our method is scalable and has the potential speed and precision necessary for large scale quantum circuit compilation.
\end{abstract}
\section{Introduction}

Compilation of quantum circuits has been investigated from different perspectives. Only recently, with the advent of NISQ devices, did compilation methods start to address optimality in the context of large scale circuits and  hardware topology constraints. One of the first works presenting a systematic method to evaluate the performance of running a circuit compiled to a particular qubit layout was \cite{isailovic2008running} -- it discusses ancilla qubit factories, interconnects and logical computation units. The quantum arithmetic as a distributed computation perspective was presented in \cite{meter2008arithmetic}. The analogy between quantum circuits and communication networks has been presented for error-corrected CNOT circuits in \cite{javadi2017optimized}. Some recent works on gate parallelism during compilation investigate how the same device can be shared for multiple circuits \cite{niu2021enabling}, and  how edge-coloring and subgraph isomorphism are related to the parallel scheduling of gates \cite{guerreschi2018two}. Organizing qubits into specialized regions has been analyzed for ancillae by \cite{square}. Exact and not scalable methods for the computation of optimal SWAP circuit depths have been introduced by \cite{wille}.

This work is motivated by the need to determine automatically, as fast and precise as possible the \emph{average} SWAP circuit depth when compiling to an arbitrary hardware layout (not necessary a regular 3D one like in the following). To the best of our knowledge, this is the first work in which the optimal SWAP depth is predicted in order to support the circuit compiler. At the same time, no work treated a complete circuit as a network of queues. We present a proof of concept and investigate the feasibility of using queuing networks -- we use the analogy between SWAP depth and input-output mean response time.

We report preliminary results after testing our approach by compiling a multiplier \cite{munoz2018quantum} to a 3D hardware layout. Most quantum devices have 1D or 2D qubit layouts, and 3D (e.g. neutral atom devices and photonic quantum technologies) is not considered viable in the short-term. However, for the purpose of this work, we chose a 3D qubit layout because we assumed that: a) routing the qubits in 3D may shorten the resulting depth; b) it is difficult for automated compilation methods; c) it is useful for developing novel compilation heuristics.

\begin{figure}
    \centering
    \includegraphics[width=0.3\columnwidth]{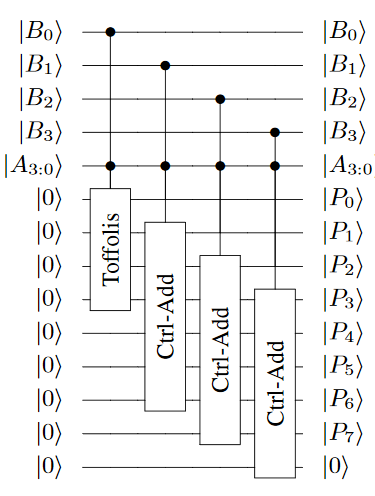}
    \caption{The multiplication circuit from \cite{munoz2018quantum} consists of three steps: 1) the Toffoli gates, 2) a sequence of controlled addition circuits; 3) the SWAP circuits occurring between every controlled addition. The third step is not illustrated.}
    \label{fig:munoz}
\end{figure}

We treat our circuit as a network of single element queues inter-connecting input and output queues. The multiplier has a highly regular structure, and we chose to compile the circuit manually. The circuit can be divided into three steps:
\begin{enumerate}
    \item a one-time application of Toffoli gates, 
    \item repeated controlled-additions on subsets of the qubit register, and
    \item setting up for the next controlled-addition step (which occurs between every controlled addition).
\end{enumerate}

The goal is to determine how the coordinates (e.g. 2D or 3D) and properties of the queues influence the compilation result. The SWAP depth of a (sub-)circuit depends not only on the qubit layout but also on the structure of the previous (sub-)circuits. We propose to use queuing network parameters (e.g. arrival rate of the queues) to capture these aspects.

In the following, in order to build the intuition behind the formal approach, we start by compiling the multiplier of Mu{\~n}oz-Coreas et. al\cite{munoz2018quantum} to a 3D lattice of qubits. Qubit queues are conveniently placed next to the adder. The queues are storing ancillae and partial product qubits. After compiling the controlled-additions to 3D, the third step of the multiplier has a constant SWAP depth of 5 gates -- irrespective of the number of qubits involved in the multiplication. All qubits were swapped in parallel without delaying (\emph{blocking}) each other.

We will not describe how we compiled the first two steps of the multiplier. It suffices to say that we used known Toffoli gate decompositions that have a 3D-like Clifford+T decomposition, and we exploited the ripple-carry structure of the controlled addition circuits.

Afterwards, we compare the prediction obtained from the queuing network analysis with our manually compiled and optimized circuit. We formalize the queuing network model of the circuit and perform a closed-form analysis to illustrate the feasibility of our approach. The analysis method has a polynomial complexity that depends on the number of network nodes. Finally, we conclude that queuing network model analysis is a promising approach towards the compilation and optimization of quantum circuits, and we formulate future work.

\section{The Multiplication Circuit}

\begin{figure}[!t]
\centering
\includegraphics[width=0.95\textwidth]{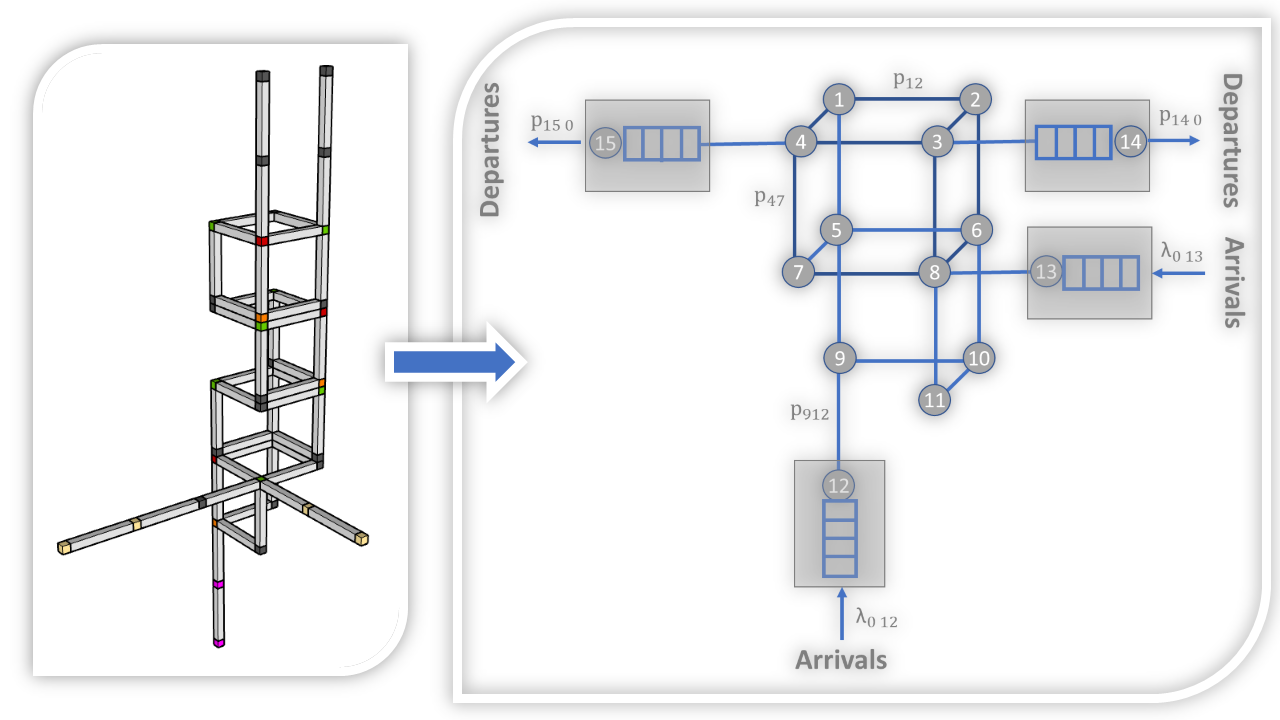}
\caption{Mapping of multiplier circuit to an open queuing network of $N=15$ nodes. LHS) The 3D qubit layout where the quantum circuit is mapped to; there are four queues (two gray, one yellow and a red one) connected to the cuboid-like 3D layout; RHS) The QN with 15 nodes, out which 4 have a finite buffer size, whereas the others (only circles with numbers 1 till 11) have a capacity of 1 (e.g. buffer size of zero). For some of the transitions, their corresponding probability $p_{ij}$ is shown. Jobs from outside arrive only at Nodes 12 and 13, and depart from Nodes 14 and 15.}
\label{fig:multiplier_QN}
\end{figure}

The structure in Fig.~\ref{fig:multiplier_QN} is designed to efficiently implement a multiplier \cite{munoz2018quantum} in a 3D nearest-neighbor environment with a minimized SWAP gate depth. By creating a single structure to implement a controlled-adder and connecting this circuit to queues within which the qubit registers are stored, the total number of qubits necessary to implement the circuit is kept small, while the SWAP depth in the third step of the multiplier has a constant depth of 5 for any $n$-bit multiplication.

There are four queues in the structure (cf. Fig.~\ref{fig:munoz} and Fig.~\ref{fig:multiplier_QN}): The top left queue, which stores used control qubits from the $B$ register, the top right queue, which stores calculated product bits from the $P$ register, the bottom-most queue which stores unused qubits from the $P$ register (all initialized to $\ket{0}$), and the queue in the final cube which stores unused control qubits from the $B$ register. At the beginning of the calculation the top two queues are empty and the bottom two queues hold their corresponding values.

Note that the $A$ register is the only register not stored in a queue; this is because all of the $A$ qubits are used for every controlled addition step, so they are stored in constant positions along the structure. Additionally, the first $n$ $P$ qubits and the first $B$ qubit begin the computation already placed throughout the structure as opposed to beginning in the queues, and at the end of the computation the last $n$ $P$ qubits and the final $B$ qubit are stored in the structure, so at no point do any queues hold every qubit in a register. It should also be noted that when a register is referred to as ‘empty’ that means it is filled with ancilla initialized to $\ket{0}$. This is so that when a qubit is added to the queue it can swap with such an ancilla which is necessary for the next step in the computation.

Gates are initially performed in the topmost cube, followed by the one below it and so on until the necessary gates have been applied to the bottom-most cube. Due to the uncompute step, gates are then applied in the opposite direction, beginning with the bottom-most cube and moving up to the top-most cube, at which point a single controlled-addition step is complete. More importantly, after each controlled-addition step a single value from both the $B$ and $P$ queues will have moved from the bottom queues into the structure, and another value from the $B$ and $P$ queues will have moved from within the structure into their corresponding top queues.

The qubits are initially positioned to implement the first step of the multiplier, then SWAP gates are applied to prepare for repeated applications of the second and third steps. After the application of the first step a single qubit from both the $P$ and $B$ registers moves forward into the structure from the bottom queues, and a single qubit from both registers also moves out from the structure into the top queues in a manner identical to the end of each controlled addition step.

For each moment in the calculation, every qubit value can swap with a single neighboring qubit. In this structure, there are three positions which only have two neighbors, $n-2$ positions with three neighbors, and every other position has four neighbors. So the majority of qubits can move in one of four different directions at each moment, or choose not to move.

\section{A Blocking Queuing Network}

To analyze the quantum multiplier circuit, we model it using a queuing network. A queuing network (QN) consists of a set of queuing systems (also called \emph{nodes} in the following), where each such system is connected to the others with some probabilities. We consider the network being open: jobs arrive from outside (\emph{source}), and after being serviced by different nodes, leave the network (\emph{sink}).

For the purpose of this work, we use a small, open QN with 15 nodes (see Fig.~\ref{fig:multiplier_QN}). We have chosen this simple structure because, as mentioned in the Introduction, we knew where our qubits will be located after a controlled addition circuit. The intermediate network which connects Nodes 1 to 11 includes nodes which can hold at any time only one job (no buffer and have a capacity of 1). Jobs from the outside arrive at two specific nodes namely Nodes 12 and 13. We assume that those two nodes have a finite buffer size. Jobs after entering the network from Nodes 12 and 13, are routed, through the nodes of the intermediate network, to the sink Nodes 14 and 15. The sink nodes, have the same characteristics in terms of buffer size as the source Nodes 12 and 13.

SWAP gate parallelism implies that qubit paths are independently running through the hardware layout, but that the paths are not blocking at intersections: a qubit does not need to wait for another qubit to cross a node (there are no bottlenecks). In order to model qubit swapping, we use a blocking queuing network. If the target nodes, to which the job needs to be transmitted after being serviced by a queuing system $i$, are full, then the job at $i$ is blocked until one of the target nodes becomes free to process this job. 

\subsection{Modeling the Network}

The open network model uses the following parameters: (1) The number of nodes of the network $N$, (2) routing (or transition) probability $p_{ij}$ of jobs from node $i$ to $j$, (3) probability $p_{0i}$ that arriving job enters the QN at node $i$, (4) probability $p_{i0}$ that job leaving node $i$ also leaves the QN such that $p_{i0}=1-\sum_{j=1}^{N}p_{ij}$, (5) arrival rate $\lambda_{0i}$ of jobs from the outside to node $i$ such that the overall arrival rate to QN is given by $\lambda=\sum_{i=1}^{N}\lambda_{0i}$. The parameters determine the QN model and, for example, the arrival rate $\lambda_i$ for each node $i$ can be derived using the traffic equation such that $\lambda_i=\lambda_{0i}+\sum_{j=1}^{N}p_{ji}\lambda_{j}$.

This mathematical formulation allows us to carry out a closed-form analysis (see below) for the steady-state probabilities. Our objective is to obtain the state probability vector ${\pi}=[\pi_1, ..., \pi_{15}]$ such that $\sum_{i=1}^{15} \pi_i=1$. Obtaining a closed-form solution for such a network consisting of 15 nodes with each node having at least 3 states (as we will see later) is not trivial: the state space size explodes to 14 million (i.e. $3^{15}$) states!

Therefore, we chose a product-form queuing network (PFQN) approach. Such networks consist of a special type of nodes only, where the underlying state-space does not have to be generated for evaluation:
$$\pi(S_1, S_2, ..., S_N)=\frac{1}{G}[\pi(S_1)*\pi(S_2)* \dots * \pi(S_N)]$$
where $G$ is a normalization constant, and $S_i$ is the specific state of node $i$. We adopt the Jackson network model for PFQN, where in order to calculate the steady-state probability of the whole network, it suffices to calculate the marginal probability of each node:
\begin{equation}
   \pi(S_1, S_2, ..., S_N)=\pi_1(S_1)*\pi_2(S_2)* \dots * \pi_2(S_N)\label{eq:1}
\end{equation}

\subsection{Modeling the Nodes}

The nodes of the circuit network are of two types: boundary and non-boundary. Boundary nodes are where jobs arrive (Nodes 12 and 13) or leave (Nodes 14 and 15) the network. Non-boundary nodes belong to the intermediate network (Nodes 1 till 11). All nodes are of the type M/M/1-FCFS: arrival and service processes are [M]arkovian (e.g. inter-arrival and service times are exponentially distributed), each node consists of one server, and jobs are processed in first-come-first-served (FCFS) fashion \cite{basmadjian}.

The difference between the boundary and non-boundary nodes is that the former have a capacity of $K$, and the latter have a capacity of one. Thus, the nodes are either M/M/1/K-FCFS or M/M/1/1-FCFS systems.

To model and analyze a node (queuing system) named $i$, the following parameters are required: (1) The different states $S_i$ that the system can have, (2) the arrival and service rates of jobs $\lambda_i$ and $\mu_i$ respectively, (3) the number of servers $m_i$, and the size of the buffer $K_i$ in case of a finite capacity queuing systems. In this paper, the considered state-space of each queuing system is discrete and the timing convention is continuous. These parameters are used to generate a continuous-time Markov chain (CTMC). The CTMC allows us to produce the square generator matrix $Q$ which presents the transition rates between two states $l$ and $m$ of the queuing system $i$ under study, such that the diagonal elements of the matrix $q_{ll}=-\sum_{m, m\neq l} q_{lm}$.

A closed-form solution (see below) is obtained by solving a set of linear global balance equations originating from $\pi_i*Q_i=0$, where $\pi_i$ is the vector of the steady-state probabilities $\pi_i(l)$ such that $\sum_l \pi_i(l)=1$ for a given queuing system $i$ and its states $l$.

\subsection{Non-boundary Nodes}

We describe the states and the transitions of the non-boundary M/M/1/1-FCFS nodes. Since these nodes have a capacity of one and only a single server, then upon arrival of a new job to this node while its server is busy, this new job cannot be buffered. To model and analyze the behavior of a job-blocking queuing system, we model the states as a two-dimensional tuple of binary values.

\begin{figure}[!t]
\centering
\includegraphics[width=0.65\textwidth]{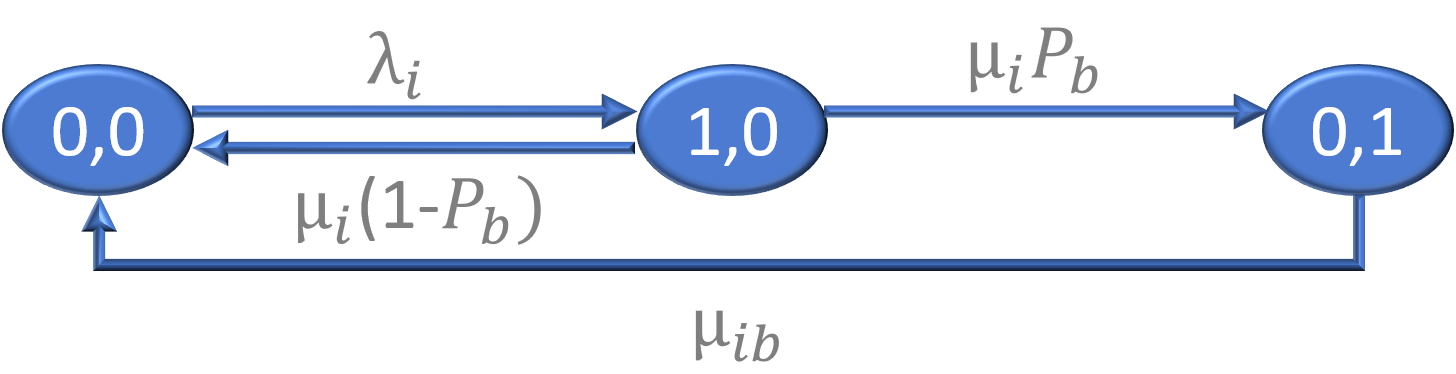}
\caption{A Markov chain for non-boundary nodes: three different states and the transitions. The first and second dimensions of each state denote the number of jobs at the service and if the job is blocked ($1/0$ = True/False).}
\label{fig:non-boundary_CTMC}
\end{figure}

Fig.~\ref{fig:non-boundary_CTMC} shows all the possible states as well as their transitions and the corresponding rates. The (0,0) is for the empty system. Upon the arrival of a new job with an arrival rate of $\lambda_i$, the system transitions from state (0,0) to (1,0). This indicates that there is a single job at the node $i$ and that the job is being serviced by a single server. After service, there are two possible transitions: (0,0) or (0,1). The former case happens when the serviced job can leave the queuing system $i$ and enter the next neighboring one $j$ in the network without being blocked by $j$. Consequently, the corresponding system $i$ becomes idle again and in (0,0) with the service rate of $\mu_i(1-P_b)$, where $P_b$ is the probability that the job after being serviced will be blocked by its neighboring node.

The transition from (1,0) to (0,1) indicates that the job after being serviced with a service rate of $\mu_i$ will be blocked by its neighbor with a probability $P_b$. The transition from (0,1) to (0,0) models the possibility of the blocked job at the queuing system $i$ to leave it and enter one of its neighboring nodes $j$. This happens because one of the queuing system $i$'s neighbors can now process the job. The queuing system $i$ becomes idle again, and this transition happens with a rate of $\mu_{ib}$ (i.e. the unblocking rate of the jobs). The steady-state probability for a full system in a finite capacity M/M/1/K-FCFS \cite{bolch} is Eq.~\ref{eq:2}. 

\begin{equation}
 \pi_i(K)=\rho_i^K\frac{1-\rho_i}{1-\rho_i^{K+1}}, \text{~where~} \rho_i=\frac{\lambda_i}{\mu_i}
 \label{eq:2}
\end{equation}

\subsection{Closed-form Analysis}

We focus on a single job running through the network, but assume that the job can take exclusive ownership of a node (see Fig.~\ref{fig:non-boundary_CTMC}). To determine the mean response time of a an arbitrary job running between source and sink, we consider that there is a non-zero probability of each node to be blocked at some point by some other job that took exclusive ownership of it.

Performing a general analysis is a complex task. We make simplifying assumptions and tune some of the parameters to the circuit we would like to compile. First, we assume that $\rho_i=1$, Eq.~\ref{eq:2}. The result is that $\pi_i(K)=\frac{1}{K+1}$, and considering that the non-boundary nodes are of type M/M/1/1-FCFS, it leads to the conclusion that the probability of any neighboring node $j$ of a another node $i$ being full (e.g. $\pi_j(1)$) is 50\%.  We can derive the worst case blocking probability $P_b$ of a job as the probability that the next hop of $j$ neighboring nodes is full: $P_b=\sum_j p_{ij}\pi_j(1)=0.5$.

Second, we have to determine the rate at which jobs arrive in the queues at nodes 12 and 13. The multiplication circuit consists of three steps (see Introduction), and herein our goal is to determine the SWAP duration during the last step. Therefore, we need to select values for $\lambda_{12}$, and $\lambda_{13}$ which reflect the structure of the previous two steps. After systematic trials necessary to select values which reflect the gate depths of the previous two steps, we arrive at $\lambda= \lambda_{12}+\lambda_{13}=0.15+0.1=0.25$.

We use the PFQN to calculate the marginal probabilities of each node in the intermediate network. We used SHARPE \cite{sharpe_tool} to calculate the steady-state probabilities from Table~\ref{table:performance_metrics} with respect to the CTMC from Fig.~\ref{fig:non-boundary_CTMC} and the parameters from Table~\ref{table:queuing_system_parameters}. The utilization $\rho_i = 1 - \pi_i(0,0)$, and the mean number of jobs $\bar K_i$ is the sum of the steady-state probabilities of (1,0) and (0,1). We observe that all the nodes are occupied for more than 79.5\% (e.g. $\pi_i(0,0)<20.5\%$) of the time, and have blocked jobs between 60\% and 69\% (e.g. $\pi_i(0,1)$) of the time.

The mean response time $\bar T_i$ is computed using Little's law \cite{little_law} which is the ratio between the mean number of jobs $\bar K_i$ to the arrival rate. The mean number of jobs in the network is $K=\frac{\sum_{i=1}^{11} \bar K_i}{11}=0.831$. We calculate a mean response time of $\bar T=\frac{K}{\lambda}=3.324$. This result confirms that our depth 5 SWAP circuit is close to the predicted optimal depth.

\section{Discussion}

Our method estimates average SWAP depths using a circuit modeled as a network of queuing systems. In a nutshell, the average SWAP depth indicates intuitively for a packet (i.e. qubit) the number of steps it takes to traverse a network between \emph{any} pair of source and destination queues. There may be additional constraints that the packet has to obey: for example, it should move between a predetermined pair of queues (i.e. this the case for the qubits that arrive and exit given queues). The value we observed in the previous section, 3.3, is for a packet that moves between \emph{any pair of queues}. The best value of 5 SWAPs is when moving between a specific pair of queues. 

We assumed that \emph{a single job traverses the network at a time}, but the fact that \emph{the nodes are blocking} seems to be a good model for multiple non-blocking SWAP qubit paths. Using the PFQN we observe that more than 3 hops are required on average to traverse from one of the source nodes to one of the sink nodes. This is not surprising and for the simple 3D layout that we have been using could have been determined by visual inspection: there are two source-sink routes, one of 5 hops and another one of 3 hops.

Our small example shows that the  arrival rates at the source queues influence the optimality of the average SWAP depth estimation. Our approach can prove valuable with respect to look-ahead compilation heuristics. Compilation speedups and cost improvements may be achieved by tuning queue parameters without being forced to consider the existing movement constraints. We showed in \cite{look-ahead_queuing} how queuing theory can be used for predicting when to start and stop T-gate distillations. Similar look-ahead scheduling techniques can be applied to the source queues (e.g. Nodes 12 and 13 in Fig. \ref{fig:multiplier_QN}).

Our procedure can be generalized. Similar approaches are used for modeling latency times and delays in communication networks \cite{balsamo2000product}. In this work we focused solely on multiplication circuits because these are building blocks of larger practical algorithms. The scale of those circuits is not a limiting factor: thousands of qubits (nodes) should be within the reach of PFQN methods. This work has been mostly to showcase and test the potential of our idea, and leave for future work the extension of our method to larger circuit instances.

\section{Conclusion}

We presented empiric evidence that a simple blocking PFQN can be used to model and predict the depth of SWAP circuits resulting during the compilation of circuits.  The closed-form analysis method has a polynomial complexity, because it is based on solving a set of linear equations for $\pi_i$ and per node there are only three states in the CTMC for intermediate network nodes.

The precision of our queuing network model is influenced by the arrival rates at the source queues. We did not model the correlations between sub-circuits, and leave this for future work. Another significant parameter is the per node probability of $(0,0)$.

Queuing networks may be a useful approach towards steering the automated compilation of very large scale quantum circuits to arbitrary (irregular) qubit layouts. Future work will focus on automatically modeling queue arrival rates and benchmarking larger and more diverse types of qubit layouts.

\begin{table}[!t]
\caption{The parameter values to compute the steady-state probabilities}
\centering
\begin{tabular}{|c|c|c|c|c|c|c|c|c|c|c|c|}
\hline
Node \# & 1 &  2 &  3 &  4 &  5 &  6 &  7
&  8&  9 &  10 &  11\\
\hline
$\lambda_i$&$0.94$ & $0.94$ & $0.936$ & $0.88$ &  $1.644$ &$1.596$ &1.02&	1.6&	1.18&	1.42&	0.86\\
\hline
$\mu_i$ &1 & 1 & 1 & 1 &  1 &1 &1&	1&	1&	1&	1\\
\hline
 $\mu_{ib}$ & 0.136& 0.136 & 0.13 & 0.144 & 0.17  & 0.142& 0.124&0.173	&0.175	&0.195	& 0.143	\\ 
 \hline
\end{tabular}
\label{table:queuing_system_parameters}
\end{table}

\begin{table}[!t]
\caption{The steady-state probabilities and the calculated performance metrics.}
\centering
\begin{tabular}{|c|c|c|c|c|c|c|c|c|c|c|c|}
\hline
Node \# & 1 &  2 &  3 &  4 &  5 &  6 &  7
&  8&  9 &  10 &  11\\
\hline
$\pi_i(0,0)$&$0.185$ & 0.185 & 0.181 & 0.203 &   0.135& 0.121&0.163&0.138	&0.18	& 0.165& 0.205	\\
\hline
$\pi_i(1,0)$ &0.174& 0.174 & 0.169 & 0.178 & 0.219  & 0.194& 0.166&0.221	&0.213	&0.234	& 0.178	\\
\hline
 $\pi_i(0,1)$ &0.641& 0.641 & 0.65 & 0.619 & 0.646  & 0.685& 0.671&0.641	&0.607	&0.601	&0.617	\\
 \hline
 $\rho_i$ &0.815&	0.815&	0.819&	0.797&	0.867&	0.8783&	0.837&	0.862&	0.82&	0.835&	0.795
	\\
 \hline
 $\bar K_i$ &0.815&	0.815&	0.819&	0.797&	0.867&	0.8783&	0.837&	0.862&	0.82&	0.835&	0.795
	\\
 \hline
 $\bar T_i$& 0.867&	0.867&	0.875&	0.906&	0.527&	0.550&	0.821&	0.539&	0.695&	0.588	&0.924\\
 \hline
\end{tabular}
\label{table:performance_metrics}
\end{table}

\bibliographystyle{splncs04}
\bibliography{_main}

\end{document}